# MetaSD: A Unified Framework for Scalable Downscaling of Meteorological Variables in Diverse Situations

Jing Hu[1], Honghu Zhang[1], Peng Zheng[1], Jialin Mu[1], Xiaomeng Huang[2]*, Xi Wu[1]*

## Author Information

## Affiliations

[1]Department of Computer Science, Chengdu University of Information Technology, Chengdu, China

[2]Department of Earth System Science, Tsinghua University, Beijing, China

*Corresponding authors. E-mssails

hxm@tsinghua.edu.cn

wuxi@cuit.edu.cn

# Abstract

Addressing complex meteorological processes at a fine spatial resolution requires substantial computational resources. To accelerate meteorological simulations, researchers have utilized neural networks to downscale meteorological variables from low-resolution simulations. Despite notable advancements, contemporary cutting-edge downscaling algorithms tailored to specific variables. Addressing meteorological variables in isolation overlooks their interconnectedness, leading to an incomplete understanding of atmospheric dynamics. Additionally, the laborious processes of data collection, annotation, and computational resources required for individual variable downscaling are significant hurdles. Given the limited versatility of existing models across different meteorological variables and their failure to account for inter-variable relationships, this paper proposes a unified downscaling approach leveraging meta-learning. This framework aims to facilitate the downscaling of diverse meteorological variables derived from various numerical models and spatiotemporal scales. Trained at variables consisted of temperature, wind, surface pressure and total precipitation from ERA5 and GFS, the proposed method can be extended to downscale convective precipitation, potential energy, height, humidity and ozone from CFS, S2S and CMIP6 at different spatiotemporal scales, which demonstrating its capability to capture the interconnections among diverse variables. Our approach represents the initial effort to create a generalized downscaling model. Experimental evidence demonstrates that the proposed model outperforms existing top downscaling methods in both quantitative and qualitative assessments.

# Introduction

Modeling weather and climate is an omnipresent necessity for science and society. It helps to enhance our understanding of future climate changes[1]. However, the intricate physical processes within the climate dynamical models pose significant numerical challenges, making it prohibitively expensive to conduct long-term high-resolution simulations, even with the current rapid advancements in computational power[2]. To address this limitation, deep learning has been employed to accelerate climate simulations by leveraging the rapid inference capabilities of neural networks.

Deep-learning-based downscaling belongs to the type of statistical downscaling that establishes the empirical relationships between historical large-scale atmospheric and local climate characteristics[3]. It falls within the data-driven paradigm that requires large amounts of data to learn relationships. Currently, most downscaling algorithms focus only on specific meteorological variables. For instance, there are downscaling algorithms for temperature[4, 5, 6, 7], precipitation[8, 9, 10, 11, 12], and wind[13, 14, 15, 16]. Although these algorithms are effective at its particular variable, González-Abad's study validates that in these univariate methods, where each meteorological variable necessitates a separate downscaling model, the preservation of cross-correlation between downscaled variables may not be ensured, potentially affecting the accuracy of subsequent forecasting processes.[17]. In other words, these methods lack a general-purpose utility for Earth system and ignore the physical connections between different meteorological variables[18, 19]. Regarding the deficiency in the generality of deep learning-based methods, Nguyen et al.[5] developed a unified model for multiple meteorological tasks within a data-driven framework. Building on a pretrained transformer-based architecture, their model can be adaptable to a breadth of climate and weather tasks after fine-tuning. Their experiments validated the superiority to a specific model in solving different Earth systems tasks such as forecasting, sub-seasonal to seasonal prediction and downscaling under a unified framework. However, their approach is focused on exploring the interconnections among various meteorological tasks, yet it overlooks the fundamental relationships between different sets of meteorological data. Moreover, the model requires retraining for application to new meteorological datasets, which limits its versatility.

Considering the constraints inherent in the generalization capabilities of contemporary deep-learning models across diverse meteorological variables, coupled with their oversight of foundational physical principles, this study aims to develop a unified downscaling approach to address these shortcomings. The trained model of the proposed method can seamlessly extend its downscaling capacity to encompass additional meteorological variables not included in its initial training set. In the Results section, it was confirmed that despite being trained on meteorological data exclusively from North America, sourced from modalities such as ERA5 and GFS, the proposed MetaSD (as illustrated in Fig. 1) exhibits notably enhanced downscaling performance when applied to data beyond the United States, including models like CFS, S2S, and CMIP6. This improvement is particularly significant when compared to the performance of state-of-the-art CNN- and GAN-based downscaling methods. In addition, these experimental findings demonstrate that MetaSD could generalize across a spectrum of downscaling variables derived from diverse numerical models and various spatiotemporal scales and indicates the ability of capturing the inherent correlation of different variables.

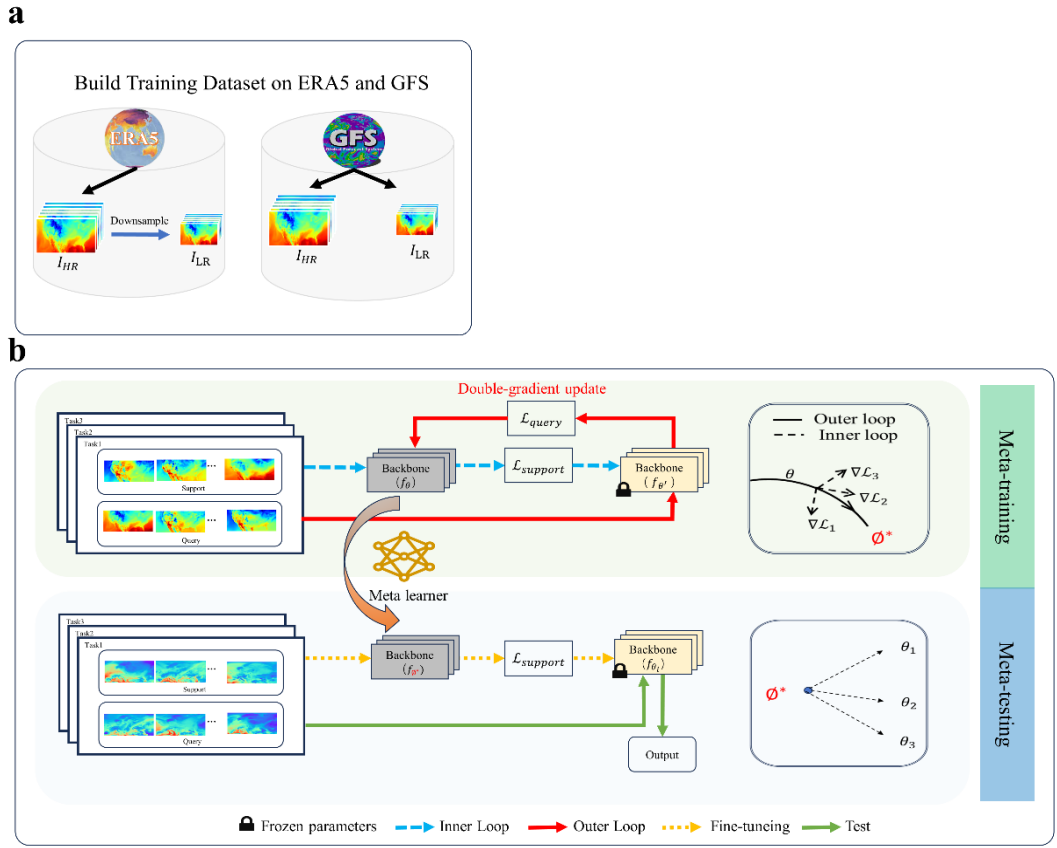

**Fig. 1 | MetaSD framework. a.** Training data are obtained from both ERA5 and GFS. Since ERA5 lacks low-resolution data, a bicubic downsampling technique is used to derive a low-resolution dataset. **b.** Meta-training and meta-testing processes in MetaSD. Meta-training is used to train a general model that can effectively adapt to various downscaling tasks[1]. During the meta-training process, the inner loop updates the model on the current task, allowing it to quickly adjust to task-specific characteristics. The outer loop updates the model by obtaining parameters that perform well on all tasks. Meta-testing is used to assess the model's generalizability on new tasks.

---

[1] In this paper, the term "downscaling task" is defined as the process of refining the resolution of a meteorological variable within a specified spatiotemporal region..

# Results

**Results of all-cross downscaling tasks using MetaSD**

To evaluate the model's generalizability, MetaSD was initially trained using variables from the European Centre for Medium-Range Weather Forecasts (ECMWF) Reanalysis V5 (ERA5) and Global Forecast System (GFS) datasets, specifically focusing on the geographical extent of the United States. The trained MetaSD was performed on variables from Climate Forecast System (CFS; part of Europe, 2022), subseasonal to seasonal (S2S; part of China, 2022), and Climate Model Intercomparison Project (CMIP6; global, 2014) data. According to Supplementary Table1, it was obvious that the testing data (E4) originate from entirely different sources than those of the training data (E0), including different variables, disparate temporal intervals, and diverse geographical locations. Besides, as shown in Supplementary Fig. 1, these testing variables have limited relevance to the training data. Such experiment was named as all-cross downscaling tasks since the testing data extends beyond the range covered by the training data.

We compared the performance of MetaSD on all-cross downscaling tasks with two interpolation baselines, bicubic[20] and kriging[21], and four state-of-the-art deep learning-based methods, DeepSD[22], EDSR[23], PhIREGAN[24] and ClimateSD[25]. Bicubic interpolation is a conventional interpolation method that performs downscaling by smoothing the data. Kriging is a geostatistical method[26] that utilizes spatial correlation among known points to predict values at unknown points. DeepSD, treating complex precipitation data as a single image, is the first method to apply deep learning to downscale meteorological data. EDSR is a deep learning-based super-resolution method that enhances image resolution by learning high-frequency details. PhIREGAN is a downscaling method based on a structure of GAN[27]. ClimateSD introduces a novel climate simulation downscaling method by using the structure of Transformer[28] and encoding the geographic information. Owing to hardware constraints, kriging, PhIREGAN, and ClimateSD are not included in our comparative experiments involving CMIP6 global data. We evaluated all methods on the basis of the peak signal-to-noise ratio (PSNR)[29], Akaikeaic information criterion(AIC)[30] and mean absolute error (MAE)[31], which are commonly used in existing deep learning-based downscaling works[22, 32]. Details of these metrics are provided in the Supplementary A. Supplementary Figs. 2-4 show a comparison of MetaSD and the baselines quantitatively, with MetaSD outperforming all the other models by achieving the highest scores across all the target variables. Upon comparative analysis, MetaSD significantly outperforms the kriging interpolation technique, evidencing enhancements of 6.62 dB in PSNR, 0.86 in MAE, and 18,105.42 in the AIC. Moreover, relative to the second-best method EDSR, MetaSD exhibits pronounced superiority, achieving notable improvements of 3.58 dB in PSNR, 0.19 in MAE, and 20,740.56 in AIC. This result indicates that despite MetaSD not being specifically trained on the testing variables, it adeptly captures the inherent interdependencies among meteorological variables within the training dataset. Consequently, this capability bolsters its efficacy in cross-downscaling tasks.

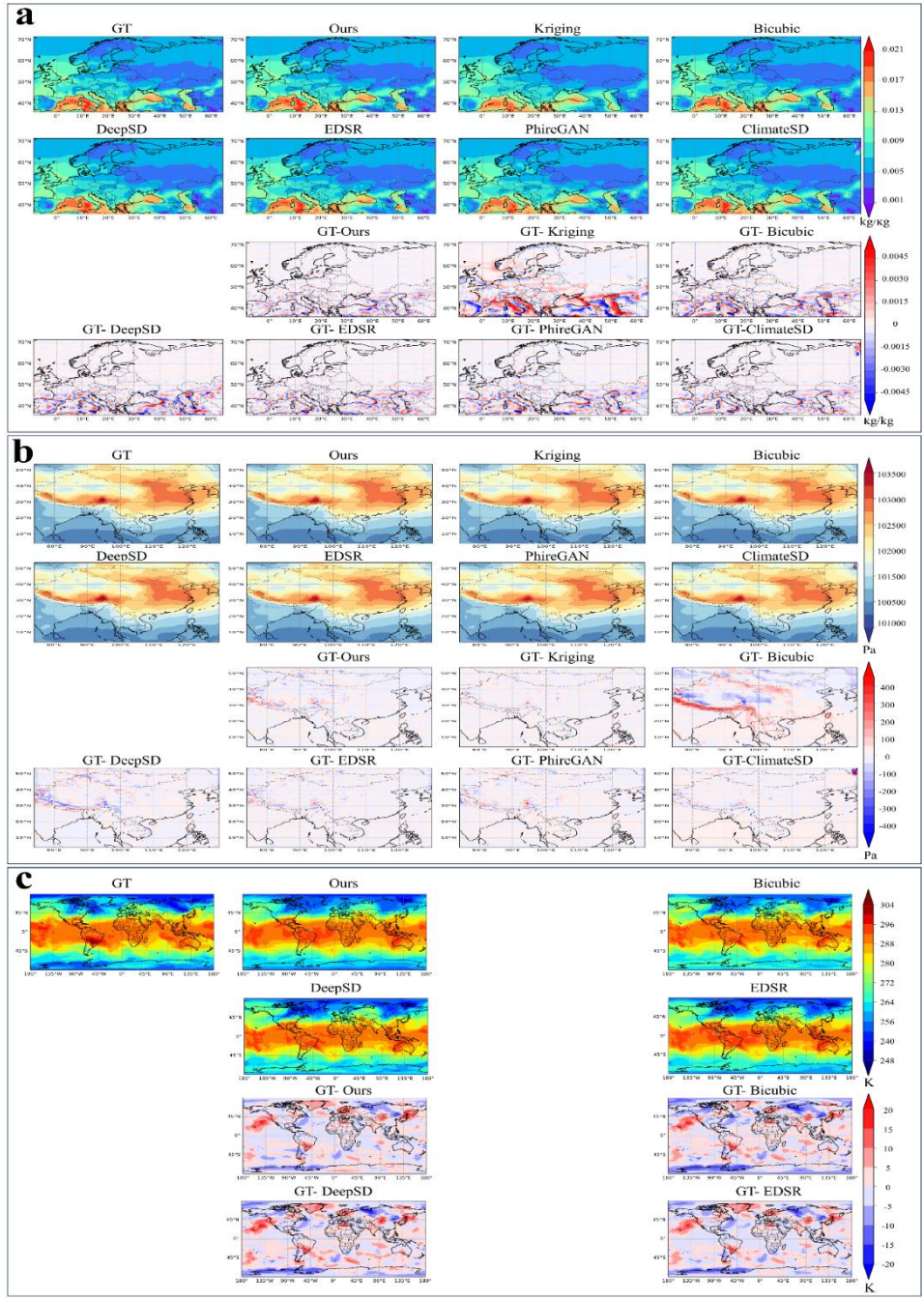

**Fig.2 | Visualization of downscaled meteorological variables from all-cross testing dataset, along with the residual map. a.** ms2 from CFS (part of Europe, 2022); **b.** msl from S2S (part of China, 2022); **c.** t850 from CMIP6(CMCC-ESM2, Global, 2014).

Figure 2 visualizes the downscaled results of the different methods, and the proposed method excels at replicating the intricate patterns of meteorological variables. In the analysis of MS2 data illustrated in Figure 2(a), all evaluated methods effectively preserved the overall distribution pattern of the data. However, MetaSD stood out for its exceptional ability to precisely identify small dry patches within largely wet areas, such as the region in the bottom left corner—a challenge where alternative approaches fell short. MetaSD not only accurately restored the general moisture levels in wet areas but also provided more intricate and true-to-reality details. A closer inspection of the residual map, MetaSD, EDSR, and ClimateSD exhibited the fewest errors among the methods evaluated, with MetaSD showing the smallest discrepancy at the transition zones between dry and wet areas. This was particularly noticeable in the case of the small island in the bottom right corner, which exhibited sharp moisture gradients. Both EDSR and ClimateSD encountered significant inaccuracies over extensive areas in this context. In contrast, MetaSD demonstrated superior performance, achieving the lowest error rate compared to all other methods tested.

The cross-downscaling results demonstrates the ability of MetaSD in learning the fundamental relationships between different sets of meteorological data. To further evaluate the robustness of MetaSD, additional three experiments were conducted using data covering a variety of geographical and time-related factors and different types of meteorological information. More specifically, the so called cross-spatiotemporal experiment is performed on E1 dataset, spanning the spatiotemporal range of China, Europe, and Australia in 1980. Compared to the suboptimal method EDSR, MetaSD is improved by 1.34 dB, 0.23, and 10,585.42 in terms of

PSNR, MAE, and AIC, respectively. For the evaluation of cross meteorological variable performance, we employ the E3 dataset, which includes factors 100m u-component of wind, 100m v-component of wind, convective precipitation, low cloud cover, skin temperature, and mean sea level pressure, and MetaSD achieves the best quantitative scores among all the compared methods. E2 dataset is used to evaluate the generalization ability to different meteorological modalities like CFSR and S2S and MetaSD still gives the best performance on this dataset, leading the second-best method by 2.77 dB in PSNR. The bar plots for the quantitative results are shown in Supplementary Figs. 6-14, and the experimental results also confirm that MetaSD effectively downsizes meteorological variables from diverse sources when compared to alternative methods.

**Comparison with a multitask learning-based downscaling framework**

The findings from cross-downscaling experiments underscore the proficiency of MetaSD in deciphering the fundamental correlations amongst various meteorological variables, highlighting its capacity to capture intricate inter-task dependencies. Concurrently, in the domain of machine learning, multitask learning (MTL) emerges as a parallel paradigm, inherently designed to facilitate the simultaneous acquisition of knowledge across related tasks[33]. This study juxtaposes the capabilities of MetaSD with an MTL-based downscaling model, with the aim of rigorously evaluating the comparative effectiveness of MetaSD in learning the underlying relationships between related tasks, thereby contributing to the broader discourse on the utility of task-agnostic learning frameworks in complex predictive modeling scenarios. To ensure fairness, we employed the same backbone architecture in the MTL-based approach and MetaSD, with the meta-learning loop (as elaborated in the Method section) being replaced by a single-layer multitask learning mechanism. Supplementary Fig. 15 illustrates the structure of MTL-based downscaling model, and the way of updating. More specifically, this model is updated by using the mean loss across all the involved tasks, which is different from MetaSD. Both models were trained on the same dataset, and they were tested on datasets E4 and E5. Supplementary Fig. 16 shows the downscaling results on E5, where the testing data within the training data domain. In assessing the comprehensive all-cross dataset E4 (detailed in Fig. 3), the MetaSD approach notably outperforms MTL-based methods, showcasing superior capability in capturing intricate details and enhancing overall fidelity and quality of the outcomes. It shows that MetaSD achieves better quantitative results to the MTL-based model, improved by 2.85dB，0.58，84,365.53 respectively in terms of PSNR, MAE and AIC. For instance, as illustrated in Fig. 3(b), the MTL-based model demonstrated a propensity for underestimation errors throughout the East Asia region. Particularly noteworthy is its response to a sudden low-pressure area in the upper right quadrant of the image, where the MTL-based model manifested significant overestimation on a large scale, in contrast to the MetaSD approach, which exhibited commendable accuracy in this context. Fig. 3(c) presented a more challenging scenario, where both models were tasked with global downscaling despite being trained exclusively on data from the United States. It is evident that MetaSD outperforms its counterpart in polar regions. Specifically, in the Antarctic, the MTL-based model was prone to marked overestimation, whereas MetaSD maintained minimal error margins. This pattern of performance was consistently mirrored in the Arctic region as well.

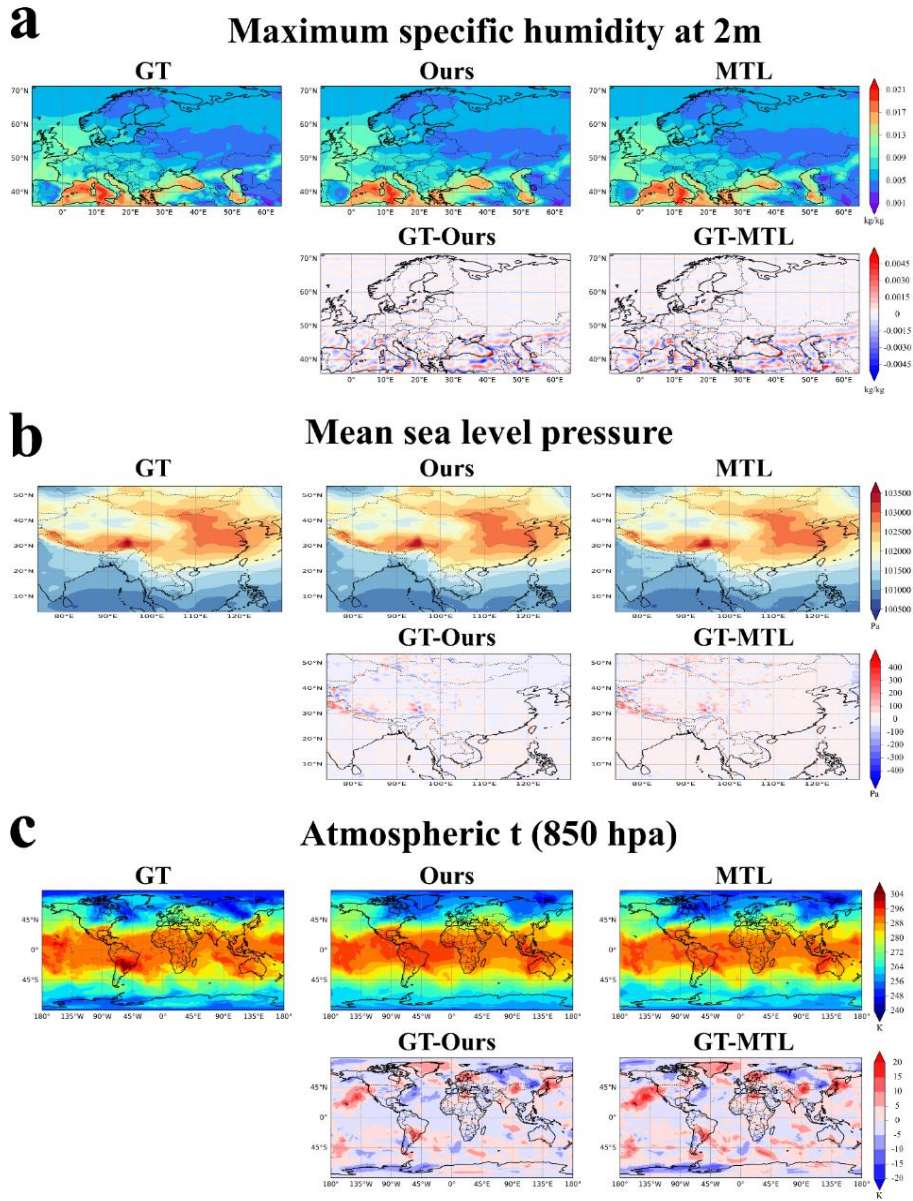

**Fig. 3 | Visual comparison images of MetaSD and MTL on the all-cross downscaling testing dataset and the residual maps. a.** ms2 in CFS (part of Europe, 2022). **b.** msl in S2S (part of China, 2022). **c.** t850 in CMIP6 (CMCC-ESM2, Global, 2014).

**The mechanism that underlies the efficacy of MetaSD**

The merit of MetaSD lies in its capability of handling downscaling tasks whose data is beyond the training set. To understand the mechanism that drives it, we utilize an analytical tool named projection-weighted canonical correlation analysis (PWCCA)[34] to determine whether MetaSD learns quickly or reuses features effectively to solve new tasks. As elaborated in Supplementary A, PWCCA offers a method for comparing the latent representation of two layers within a neural network, yielding a similarity score ranging from 0 (indicating no similarity) to 1 (denoting identical representations). When a meta-learning model benefits from rapid learning, substantial adjustments occur in both feature representation and neural network parameters as it adapts to new tasks, resulting in a reduced PWCCA score. The adjustments induced by rapid learning were primarily influenced by the weight conditioning established during the initial meta-learning phase. This foundational process enables the model to promptly assimilate new information and dynamically reorganize its internal structure to meet evolving challenges Conversely, when a meta-learning model demonstrates an advantage of versatility through feature reuse, the meta-initialization phase would successfully embed highly relevant features within the model, resulting in a high PWCCA score. This indicates that the model's initial setup is robust and contains a wealth of information, enabling effective transfer and application to a broad range of subsequent tasks without significant reconfiguration [35].

Leveraging the architecture of the well-known meta-learning framework MAML[36], MetaSD incorporates a dual optimization loop structure, comprising an outer loop dedicated to identifying a suitable meta-initialization and an inner loop designed for rapid adaptation to novel tasks[35]. To explore the mechanism of MetaSD's generalization capabilities, our investigation focused on the inner loop's dynamics. We meticulously calculated the PWCCA score for each layer of within MetaSD, both prior to and subsequent to the execution of the inner loop process. These PWCCA scores are visually presented in Fig. 4. In the first three layers of the neural network, the correlation coefficients of PWCCA are all higher than 0.9. This indicates that the effect of the shallow network does not change significantly before and after the inner loop adaptation. To visually underscore these disparities, Fig. 5 provides visual representations of feature maps from two distinct layer types throughout the inner loop, specifically highlighting a consistent feature map in the shallow layer (layer 1) and notable variations in the deep layer (layer 14). According to the feature maps of layer 1, where there is virtually no change of PWCCA before and after adaptation, further illustrating that the shallow network primarily utilizes previously learned network features. As the number of network layers increases, the PWCCA correlation coefficients significantly decrease, with some tasks even dropping below 0.5. This suggests a substantial gap between the current new task and the tasks the model was trained on, requiring the deeper layers of the network to not merely rely on previously learned features but to quickly adjust themselves to adapt to new tasks. The feature maps of layer 14 shows a noticeable change of PWCCA values before and after adaptation. Evidently, the adaptation process in MetaSD brings about minimal changes in its shallow layers while inducing substantial transformations in the deeper layers, which follows an established principle is that early neural network features tend to be more generalized, evolving toward a task-specific nature as one delves deeper into the network[37]. We argue that the main benefit of meta-learning in the meteorological domain is not due to feature reuse but rather rapid adaptation. Nevertheless, it should be noted that a very small number of variables, such as the meteorological variable cape, also have high PWCCA values in the deeper layers. This is because it is very similar to the meteorological variable (tp) used in the training set, thus the network parameters do not need to be adjusted too much to quickly adapt to the downscaling task of cape.

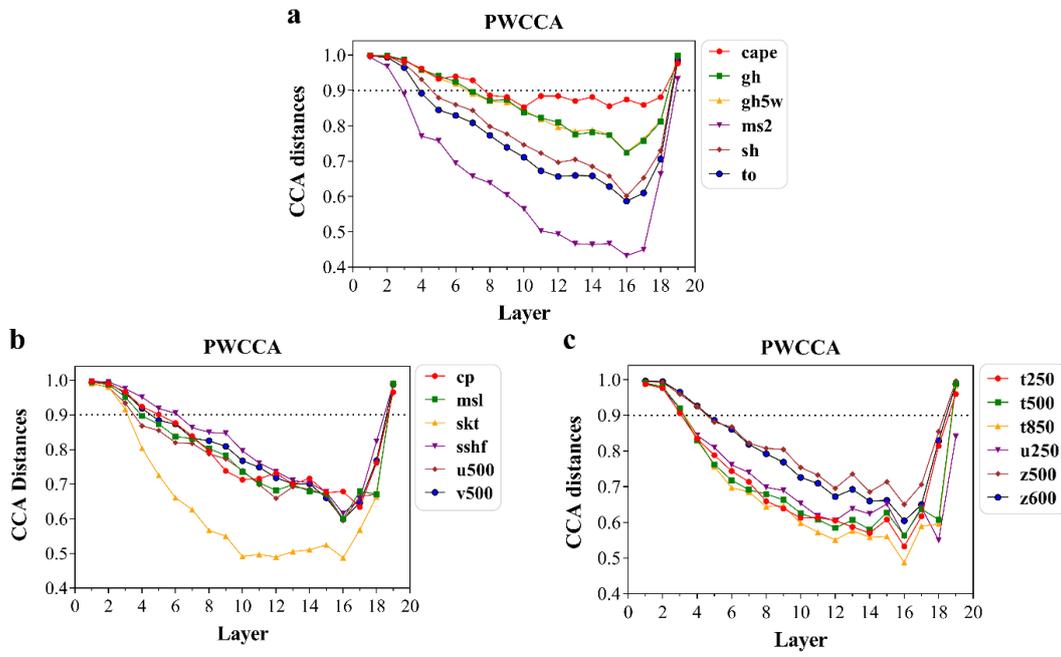

**Fig. 4 | The similarity score, calculated by PWCCA, of each layer in MetaSD before and after the inner loop procedure. a.** Meteorological variables in CFS mode (part of Europe, 2022) **b.** Meteorological variables in the S2S mode (part of China, 2022). **c.** Meteorological variables in CMIP6 mode (Global, 2014)).

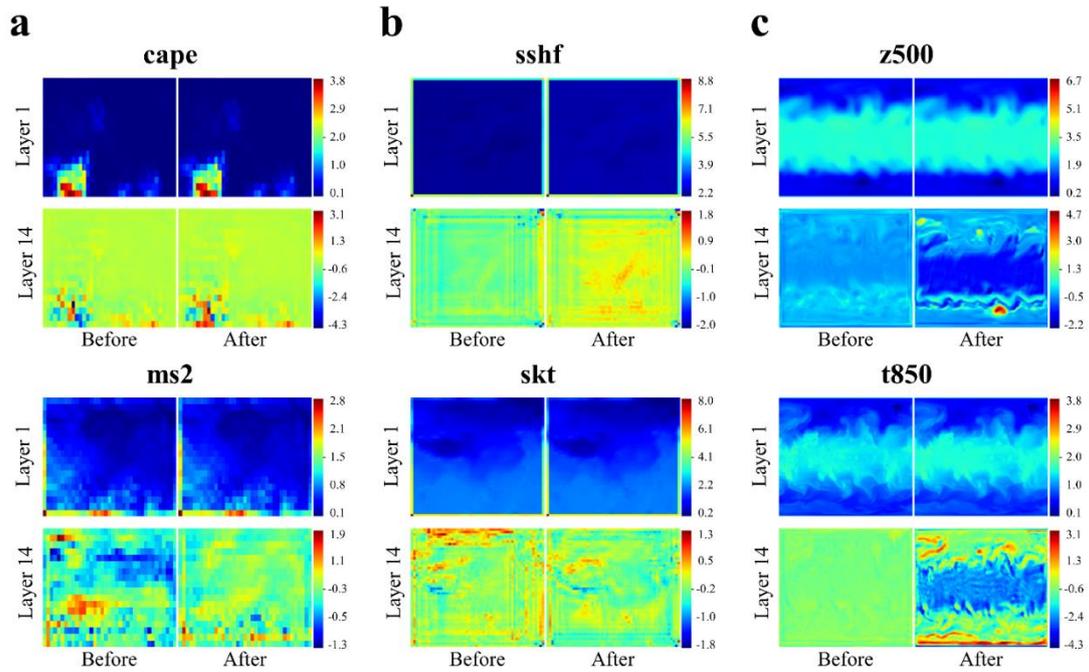

**Fig. 5 | Feature maps of the shallow layer (layer 1) and deep layer (layer 14) before and after inner loop adaptation. a.** cape and ms2 in CFS (part of Europe, 2022). **b.** sshf and skt in S2S (part of China, 2022). **c.** z500 and t850 in CMIP6 (CMCC-ESM2, Global, 2014).

## Discussion

Recent developments in the field of deep learning have progressively introduced novel "universal" models, which shave been applied in several computer vision and natural language processing fields[38], such as segmentation[39], speech recognition[40], classification and object detection[41]. Meanwhile, contemporary investigations on incorporating deep learning with meteorology and climatology still overspecialize on a single task, thus falling short in terms of versatile applicability spanning a broad variety of tasks[32]. Although foundational models such as ClimaX[32] and Pangu-Weather[42] open up new opportunities for general models in the field of climatology, all these approaches rely on a pretrained large language model architecture, necessitating training on extensive datasets.

This paper focuses on climate downscaling as crucial for obtaining high-resolution climatic data in specific geographic areas. We advocate for a lightweight meta-learning approach over reliance on large language models, showing its effectiveness in revealing intricate relationships among meteorological variables and its ability to generalize to unseen scenarios. Our proposed MetaSD model outperforms other deep-learning-based downscaling techniques across diverse experiments, showing superior performance in various geographical regions, timeframes, and data types. MetaSD's generalization ability stems from its rapid adaptation to new tasks and its identification of optimal initialization parameters with sensitivity and transferability across different downscaling tasks. Particularly, MetaSD exhibits significantly lower AIC values, suggesting that it achieves an enhanced balance between precisely producing high-resolution data and minimizing model complexity, as measured by the model parameter number. Moreover, when directly compared to a MTL-based model, MetaSD consistently demonstrates superior performance. The key disparity between MetaSD and MTL lies in their optimization strategies: MetaSD employs meta-learning, which optimizes the model across a vast array of future tasks sampled from an unknown distribution, whereas the MTL approach emphasizes cotraining, optimizing the average model over a finite set of known tasks [43].

Although MetaSD consistently performed well in various scenarios, it fell short of achieving the highest quantitative results in certain downscaling tasks. For instance, in the all-cross experiment, Kriging outperformed MetaSD for the sshf factor, while EDSR excelled for the v10 factor in cross-modality experiments. Moreover, in comparisons with MTL, MTL exhibited superior performance for the cp factor in all-cross experiments. These findings suggest the downscaled tasks used for training are not sufficiently diverse to represent the distribution of meteorological downscales as a whole. Future advancements may involve choosing the representative samples. In addition, MetaSD is limited to fixed-magnitude downscaling and does not address arbitrary-magnitude or spatio or temporal downscaling. Expanding its capabilities to more flexible downscaling magnitude presents a significant future challenge in enhancing meteorological downscaling frameworks.

# Methods

## Datasets

This study utilized six distinct and widely used meteorological modalities, each offering unique capabilities and data sources. The ERA5 reanalysis data[44], with a 0.25° × 0.25° global grid resolution, integrate the state-of-the-art IFS[45] of the ECMWF. The GFS (http://www.emc.ncep.noaa.gov/GFS) is a global model developed by the National Center for Environmental Prediction that runs four times a day at three grid resolutions (0.25° × 0.25°, 0.5° × 0.5° and 1° × 1°). The CFS[46] and its reanalysis counterpart CFSR[47] encapsulate comprehensive Earth system interactions. The CMIP6 dataset[48] 1provides an extensive collection of climate model results, facilitating the evaluation of future climate projections under various scenarios. Finally, the S2S Prediction Project[49] represents an international initiative aimed at improving the accuracy and reliability of weather and climate forecasts on subseasonal to seasonal timescales, encompassing periods ranging from days to weeks and months.

The training dataset, incorporating data from ERA5 and GFS, includes regional information specific to the U.S., covering a latitudinal range of 25°N to 48.75°N and a longitudinal range of 70.25°W to 130°W, sourced from various years, including 2000 and 2020 for ERA5 and 2023 for GFS. The high-resolution (HR) dataset, termed as $\mathbf{I}_{HR}$ featuring a grid size of 96×240, is a direct output of climate simulation models, offering a spatial resolution of approximately 25 kilometers or 0.25 degrees, while the corresponding low-resolution (LR) data, termed as $\mathbf{I}_{LR}$ with a grid size of 24×60, were created through either bicubic interpolation from the HR output, resulting in a coarser resolution of approximately 100 kilometers, or by utilizing climate simulation model output from a coarser 1-degree grid, as is the case with GFS.

Five different testing datasets were generated: E1, known as the cross-spatiotemporal dataset, contains variables sourced from ERA5, covering regions in China, Europe, and Australia for the year 1980; E2, termed the cross-mode dataset, encompasses variables extracted from CFSR and S2S for the United States in 2000; E3, as the cross-variable dataset, consists of distinct meteorological variables sourced from ERA5 in the United States for the year 2020; E4, designated as the all-cross dataset, comprises variables obtained from CFS, S2S in 2022, and CMIP6 in 2014; and E5 contains data that fall within the same geographical and temporal scope as the training dataset. Comprehensive details of all the data employed in the experiments can be found in Supplementary Table 1.

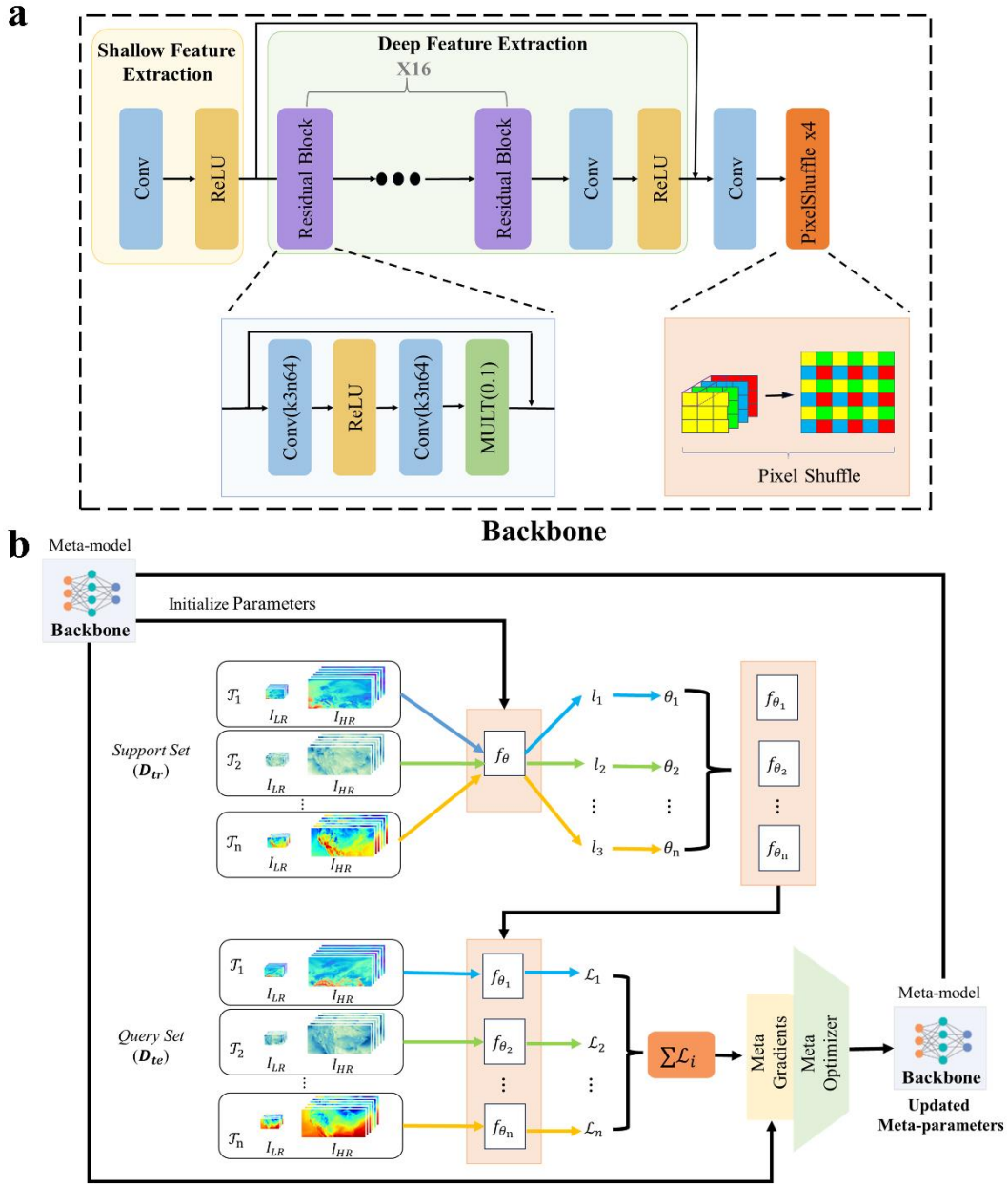

**Fig. 6 The MetaSD framework. a.** Structure of the backbone used for downscaling. **b.** Schematic diagram of the meta-learning framework (MetL).

## Neural Network of MetaSD

The MetaSD framework comprises two fundamental components: the downscaling network (backbone, illustrated in Fig. 6(a)) and the meta-learning framework (MetL, depicted in Fig.

6(b)). The backbone is responsible for establishing the mapping relationship between images of various scales within the downscaling task, and for the sake of simplicity, the structure of the EDSR method is adopted[23]. MetL, on the other hand, functions as a meta-learning module, extensively exploring the inherent relationships among diverse climate downscaling tasks, thereby enhancing collaborative learning across a wide range of downscaling tasks

Climate variables, climate data sources, and spatial and temporal ranges are the four factors that influence the downscaling task. In MetaSD, each variation in these factors is treated as a distinct downscaling task. For convenience, the distribution of the downscaling tasks is termed as $\rho(T)$. The training set of MetaSD is generated by sampling $t$ different tasks ($\{T_i\}_{i=1}^{t}$) from $\rho(T)$. For each task, $N$ pairs of $\{\mathbf{I}_{LR\_i}, \mathbf{I}_{HR\_i}\}$ were randomly selected. Among these pairs, $\lfloor N/2 \rfloor$ pairs comprised the support set ($D_{tr}$) used for quickly adapting to new downscaling tasks in the inner loop. The remaining $\lfloor N/2 \rfloor$ pairs of images form the query set ($D_{te}$), which is employed to find an initial parameter that can better generalize to various downscale tasks.

In MetaSD, we utilize the $l_1$ loss as the loss function, as depicted in Eq. (1).

$$L(\theta) = E_{(I_{HR}, I_{LR})} \left[ \| I_{HR} - f_\theta(I_{LR}) \|_1 \right] \tag{1}$$

where $f_\theta(\cdot)$ represents the backbone structure.

The meta-learning module of MetaSD encompasses two loops. The objective of the inner loop is to enable the meta-model to swiftly adjust to new downscaling tasks, that is, to adapt to the characteristics of a new task within a specified maximum number of steps, such as within 30 training steps. To achieve this goal, a small batch of data (task batch) is sampled from $D_{tr}$ to further fine-tune the meta-model's parameters ($\theta$ in Eq. 1), which are tailored to the selected downscaling task $T_i$. In detail, to adapt the meta-model to the specific task $T_i$, the meta-model's parameters $\theta$ in $f_\theta(\cdot)$ are updated as $\theta_i$ using three iterations of stochastic gradient descent (SGD), as described in Eq. (2). This adaptation aims to enable the meta-model to quickly converge to an optimal set of parameters for that task:

$$\theta_i = \theta - \alpha \nabla_\theta L_{T_i}^{D_{tr}}(f_\theta) \tag{2}$$

where $\alpha$ is the step size for task-level training.

The outer loop of MetaSD is responsible for optimizing the initial parameters across multiple downscaling tasks. The parameters of MetaSD are trained by optimizing the average performance of $f_{\theta_i}$ with respect to $\theta$ across tasks sampled from $\rho(T)$ within a meta-batch ($B_{meta}$ in the following text). However, this approach can potentially lead to a problem known as meta-overfitting or meta-generalization[50]. Specifically, meta-overfitting occurs when the model parameters are overly optimized for the specific tasks in the training set and fail to generalize well to new, unseen tasks. In the context of downscaling, the inclusion of both easy and challenging tasks within a unified $B_{meta}$ may inadvertently divert attention from more intricate tasks, leading to potential overfitting of the meta-model to simpler tasks. To address

the above issue, a weighted parameter update approach for the outer loop of MetaSD was designed as follows:

$$x_i = \min\left(0.27 + \frac{n \cdot L_{T_i}^{D_{te}}(f_{\theta_i})}{2 \cdot \sum_{B_{meta}} L_{T_i}^{D_{te}}(f_{\theta_i}) + \varepsilon}, \ 0.99\right), \varepsilon \to 0^+$$
(3)

$$w_i = \max(-x_i^3 \cdot \log(1 - x_i), 1)$$
(4)

where $x_i$ ranges from $(0.27, 0.99]$, $n$ represents the total number of tasks, $w_i$ denotes the loss weight of the $i$-th task. The loss weight is higher for difficult tasks and lower for easier tasks. By increasing the contribution of challenging tasks through loss weighting, the meta-model is prevented from overfitting to easy tasks.

In each iteration of the outer loop updating, the model utilizes the temporary $\theta_i$ parameters from the inner loop to evaluate its performance on the query set of each task. This performance is subsequently used to update the initial model parameters, aiming to optimize the generalizability of the model, enabling the model to better generalize across various downscaling tasks. Specifically, we optimize the model parameters $\theta$ to minimize the test error of task $D_{te}$ relative to the weights $\theta_i$ sampled from task distribution $\rho(T)$. The optimization objective is shown in Eqs. 5 and 6.

$$\arg\min_{\theta} \sum_{B_{meta}} L_{T_i}^{D_{te}}(f_{\theta_i}) = \arg\min_{\theta} \sum_{B_{meta}} w_i L_{T_i}^{D_{te}}\left(\theta - \alpha \nabla_{\theta} L_{T_i}^{D_{tr}}(f_{\theta})\right)$$
(5)

$$\theta \leftarrow \theta - \beta \nabla_{\theta} \sum_{B_{meta}} w_i L_{T_i}^{D_{te}}(f_{\theta_i})$$
(6)

where $\beta$ is the meta step size.

During the testing phase, we select a new downscaling task as the testing task and utilize the model parameters learned during the meta-training phase as the initial parameters. Next, we input the training samples of the testing task into the model and perform a small amount of parameter fine-tuning to quickly adapt to the specific characteristics of the current testing task. Subsequently, we generate predicted high-resolution results on the testing samples of the task and evaluate the downscaling performance of the model by comparing the results with the ground truth using metrics such as the Akaike information criterion.

Training details

The complete algorithm is detailed below. This process was carried out on an Ubuntu 18.04 system, including a Pascal Tesla P100 GPU graphics processing unit (GPU) with a video memory size of 16 GB. Three thousand LR-HR pairs per downscaling task were utilized, and the model was trained using the L1 loss function and the adaptive moment estimation (Adam) optimizer. A learning rate of 0.01 was used in the inner loop, which was set to be large to optimize the parameters for the generalized downscaling task in a few steps. This process involved 3 gradient updates and 3 unrolling steps to obtain the adapted parameters. For the outer loop, a learning rate of 0.0001 was chosen to optimize the model parameters. This rate is set to be low because the model needs to slowly converge on the most appropriate parameters for all downscaling tasks after numerous iterations. In the fine-tuning phase, the initial learning rate was set to 1e-2, a mini-batch size of 20 was used, and the number of epochs was set to 30.

**Algorithm:** Meta-learning framework (MetaSD)

**Input:** Learning rates: $\alpha$, $\beta$; distribution over tasks: $\rho(T)$;

**Output:** Model parameter $\theta_M$

1:   Randomly initialize $\theta$
2:   **while** not done **do**
3:       Sample batch of tasks $T_i \sim \rho(T)$
4:     **for** all $T_i$ **do**
5:         Sample $k$ pairs examples $D_{tr} = \{I_{LR\_i}, I_{HR\_i}\}$ from $T_i$
6:         Calcualte the loss $L_{T_i}^{D_{tr}}(f_\theta)$ according to Eq. (1)
7:         Compute adapted parameters with gradient descent: $\theta_i = \theta - \alpha \nabla_\theta L_{T_i}^{D_{tr}}(f_\theta)$
8:         Sample $k$ pairs examples $D_{te} = \{I_{LR\_i}, I_{HR\_i}\}$ from $T_i$ for the meta-update
9:     **end for**
10:    Calculate task weights $w_i$ by Eqs. (3) and (4)
11:    Update $\theta$ with respect to average weighted loss: $\theta \leftarrow \theta - \beta \nabla_\theta \sum_{B_{meta}} w_i L_{T_i}^{D_{te}}(f_{\theta_i})$
12: **end while**

# Data Availability

The data used in the experiment were obtained from publicly available datasets and can be accessed through the following link.

ERA5 data are available at https://cds.climate.copernicus.eu/cdsapp#!/dataset/reanalysis-era5-single-levels?tab=form and https://cds.climate.copernicus.eu/cdsapp#!/dataset/reanalysis-era5-pressure-levels?tab=form.

GFS data are available at https://nomads.ncep.noaa.gov.

CFS data are available at https://www.ncei.noaa.gov/data/climate-forecast-system/access/operational-analysis/6-hourly-by-pressure.

CFSR data are available at https://www.ncei.noaa.gov/products/weather-climate-models/climate-forecast-system.

S2S data are available at https://apps.ecmwf.int/datasets/data/s2s/levtype=sfc/type=cf. CMIP6 data are available at https://aims2.llnl.gov/search.

## Code Availability

Our code is available at https://github.com/xyxy880/MetaSD.

# Acknowledgements

This study is supported by the following grants: National Key Research and Development Program of China (2020YFA0608000), National Science Foundation of China (42075142, 42375148, 42125503, 42130608), Sichuan Science and Technology Program (2023YFG0305, 2022YFG0042, 2023YFG0025), Opening Foundation of Agile and Intelligent Computing Key Laboratory of Sichuan Province, CUIT Science and Technology Innovation Capacity Enhancement Program project under Grant KYQN202305.

## Author Contributions

J.H. proposed the idea and designed the project. H.Z. developed the methods and performed the benchmark. P.Z. prepared the data, performed the benchmark and analyzed the results. J.M. contributed to the interpreting results, discussions of associated dynamics and improvement of the presentation. X.W. and X.H. supervised the work. All authors wrote the manuscript and approved the final version.

## Ethics declarations

The authors declare no competing interests.